\documentclass[prd,aps,twocolumn,superscriptaddress,showpacs,dvips]{revtex4}
\usepackage{feynmp}
\usepackage{amssymb}
\usepackage{epsfig}
\usepackage{graphicx}
\usepackage{subfigure}
\usepackage{hyperref}

\begin{document}

\title{Confinement induced by fermion damping in three-dimensional QED}

\author{Jing Wang}
\affiliation{Department of Modern Physics, University of Science and
Technology of China, Hefei, Anhui, 230026, P.R. China}
\author{Jing-Rong Wang}
\affiliation{Department of Modern Physics, University of Science and
Technology of China, Hefei, Anhui, 230026, P.R. China}
\author{Wei Li}
\affiliation{Department of Modern Physics, University of Science and
Technology of China, Hefei, Anhui, 230026, P.R. China}
\author{Guo-Zhu Liu}
\affiliation{Department of Modern Physics, University of Science and
Technology of China, Hefei, Anhui, 230026, P.R. China}
\affiliation{Institut f$\ddot{u}$r Theoretische Physik, Freie
Universit$\ddot{a}$t Berlin, Arnimallee 14, D-14195 Berlin, Germany}

\begin{abstract}
The three-dimensional non-compact QED is known to exhibit weak
confinement when fermions acquire a finite mass via the mechanism of
dynamical chiral symmetry breaking. In this paper, we study the
effect of fermion damping caused by elastic scattering on the
classical potential between fermions. By calculating the vacuum
polarization function that incorporates the fermion damping effect,
we show that fermion damping can induce a weak confinement even when
the fermions are massless and the chiral symmetry is not broken.
\end{abstract}

\pacs{11.10.Kk, 12.20.Ds, 11.30.Rd}

\maketitle


There are generally two basic reasons why quantum electrodynamics in
three space-time dimensions (QED$_3$) is interesting and deserves
systematical investigation. First, it can exhibit dynamical chiral
symmetry breaking \cite{Pisarski, Appelquist86, Appelquist88, Nash,
Dagotto, Atkinson90, Dorey, Burden, Maris, Gusynin, Fischer,
Roberts} and confinement \cite{Burden, Maris}, which are well-known
to be two salient non-perturbative features of QCD, but has
relatively simple structure. Second, it has proved to be the
effective theory of a number of planar condensed matter systems,
including high temperature superconductor \cite{Affleck, Kim, Wen,
FranzTesa, Liu, Liu05} and spin-1/2 Kagome spin liquid \cite{Ran}.

Permanent confinement is an important concept in modern particle
physics. It was originally proposed to understand why free quarks
are invisible in all available experiments although they are widely
believed to be the elementary blocks of hadrons. While it is still
unclear how to define and describe quark confinement in the
framework of QCD, in the past thirty years there have been important
progresses in understanding confinement in three-dimensional QED,
which may shed some light on the study of permanent confinement in
QCD.

There are generally two kinds of QED$_3$: compact one and
non-compact one. In compact QED$_3$, the gauge field $A_\mu$ is an
angular variable defined on a circle. As found by Polyakov
\cite{Polyakov}, there are topological instanton solutions in the
classical equation of gauge field in this theory. He also calculated
the classical potential between static charges using the Wilson loop
criterion and revealed that the potential energy grows linearly with
distance \cite{Polyakov}, which implies that all static charges are
permanently confined. In the non-compact QED theory, however, the
gauge field $A_\mu$ is defined in the domain $(-\infty, +\infty)$.
In this case, there are no instanton excitations and thus the
arguments of Polyakov are no longer valid.

However, it is still meaningful to study the issue of fermion
confinement in non-compact QED$_3$ even though there are no
instanton configurations. According to the theoretical analysis of
Burden \emph{et} \emph{al.} \cite{Burden}, the classical potential
between fermions is primarily determined by the asymptotic behavior
of vacuum polarization function $\Pi(\mathbf{q})$ in the
$\mathbf{q}\rightarrow 0$ limit. Shortly after their work, Maris
\cite{Maris} found that the massless fermions are always deconfined
but will be confined by a logarithmic potential once they acquire a
finite mass via chiral symmetry breaking.

However, dynamical mass generation may not be the only mechanism to
induce fermion confinement in non-compact QED$_3$. In this paper, we
will study the effect of fermion damping on the classical potential
between massless fermions. Normally, in any realistic many-particle
system that can be described by QED$_3$, there are always some kinds
of disordered potential, caused by impurity atoms, defects, or other
sources. Due to the scattering by such disordered potentials, the
fermions develop a finite damping rate $\Gamma$. The fermion damping
rate has completely different physical meaning from fermion mass,
and does not break the chiral symmetry. By calculating the vacuum
polarization function and classical potential, we show that a finite
fermion damping can induce weak confinement of massless fermions.

The Lagrangian for QED$_3$ with $N$ flavors of massless fermions is
given by
\begin{eqnarray}
\mathcal{L} = -\frac{1}{4}F_{\mu\nu}^2 + \sum^N_{i=1}\bar{\psi}_i
(i\partial \!\!\!/-e A\!\!\!/)\psi_i. \label{lagrangian}
\end{eqnarray}
Here, the $4\times 4$ $\gamma$ matrices can be defined as
$\gamma_\mu = (\sigma_3,i\sigma_1,i\sigma_2) \otimes \sigma_3$,
which satisfy the standard Clifford algebra $\{\gamma_\mu,
\gamma_\nu\} = 2g_{\mu\nu}$ with metric $g_{\mu\nu} = (1,-1,-1)$.
The conjugate spinor field is defined as
$\bar{\psi}=\psi^{\dagger}\gamma_0$. This model respects a
continuous chiral symmetry $\psi \rightarrow
e^{i\theta\gamma_5}\psi$, with $\gamma_5$ anti-commutating with
$\gamma_{\mu}$.

In the Euclidean space, the free fermion propagator is
\begin{eqnarray}
S_0(p) = \frac{1}{p\!\!\!/},
\end{eqnarray}
and the free gauge field propagator is
\begin{eqnarray}
D_{\mu\nu}^0(q) = \frac{1}{q^2}\left(g_{\mu\nu} - \frac{q_\mu
q_\nu}{q^2}\right)
\end{eqnarray}
in the Landau gauge. The classical potential between fermions can be
defined in the coordinate space as \cite{Burden}:
\begin{eqnarray}
V(\mathbf{r}) = -e^2 \int\frac{d^2 \mathbf{k}}{(2\pi)^2}
e^{i\mathbf{k}\cdot \mathbf{r}}\frac{1}{\mathbf{k}^2}.
\end{eqnarray}
After direct calculation, it is easy to get
\begin{equation}
V(\mathbf{r}) \propto \ln r.
\end{equation}
This is a weak, logarithmic confining potential caused by gauge
force. However, this result should not be considered seriously
because it may be completely modified by higher order corrections.
When studying a quantum field theory like QED$_3$, it is necessary
to include the vacuum polarization function to the gauge field
propagator \cite{Appelquist86}.

With the help of Dyson equation, the full propagator of gauge field
becomes
\begin{eqnarray}
\label{full photo pragator} D_{\mu\nu}(q) =
\frac{1}{q^2[1+\Pi(q)]}\left(g_{\mu\nu} - \frac{q_\mu
q_\nu}{q^2}\right),
\end{eqnarray}
where the vacuum polarization function $\Pi(q)$ is related to the
polarization tensor by
\begin{equation}
\Pi_{\mu\nu}(q) = \left(q^2 g_{\mu\nu} - q_\mu q_\nu\right)\Pi(q).
\end{equation}
To the leading order of $1/N$ expansion, the vacuum polarization
tensor is defined as
\begin{eqnarray}
\Pi_{\mu \nu}(q) = -\alpha\int \frac{d^3 k}{(2\pi)^3}
\frac{\mathrm{Tr}[\gamma_\mu k\!\!\!/\gamma_\nu
(q\!\!\!/+k\!\!\!/)]}{k^2(q+k)^2},
\end{eqnarray}
where the dimensionless parameter $\alpha = Ne^2$.

After taking the polarization function into account, the classical
potential becomes
\begin{eqnarray}\label{potential}
V(\mathbf{r}) = -e^2\int \frac{d^2 \mathbf{k}}{(2\pi)^2}
e^{i\mathbf{k} \cdot
\mathbf{r}}\frac{1}{\mathbf{k}^2[1+\Pi(\mathbf{k})]}.
\end{eqnarray}
After straightforward calculation, Burden \emph{et} \emph{al.} found
that the potential behaves as \cite{Burden}
\begin{eqnarray}
V(\mathbf{r})\propto \frac{1}{1+\Pi(0)}\ln r + O(\frac{1}{r}).
\end{eqnarray}
Now we see that the behavior of polarization function at zero
momentum $\Pi(0)$ plays a crucial role. When the fermions are
massless, it is easy to obtain that $\Pi(q) = \frac{\alpha}{8q}$. As
$q\rightarrow 0$, $\Pi(0) \rightarrow \infty$, so the logarithmic
potential is suppressed. For a confining logarithmic potential to be
present, $\Pi(0)$ should take a finite value.

Since the pioneering work of Appelquist \emph{et} \emph{al.}, it has
been known that massless fermions can acquire a finite mass via the
non-perturbative mechanism of dynamical chiral symmetry breaking
when the fermion flavor $N$ is less than certain critical value
\cite{Appelquist88, Nash, Dagotto, Atkinson90, Dorey, Burden, Maris,
Gusynin, Fischer, Roberts}. Once the fermions become massive, the
polarization function becomes
\begin{eqnarray}
\Pi(q^{2}) = \frac{\alpha}{4\pi}\left(\frac{2m}{q^{2}} + \frac{q^{2}
- 4m^{2}}{q^3}\arcsin\frac{q}{\sqrt{q^{2}+4m^{2}}}\right),
\end{eqnarray}
where fermion mass $m$ is constant. As $q\rightarrow 0$, we have
\begin{eqnarray}
\Pi(0) \rightarrow \frac{\alpha}{6\pi m}. \label{pi_0 with a mass}
\end{eqnarray}
Hence, once the fermions acquire a finite mass via the mechanism of
dynamical chiral symmetry breaking, the fermions are then confined
by a logarithmic potential \cite{Maris}.

It is interesting to ask a question: Can fermion confinement be
induced by other mechanisms than dynamical chiral symmetry breaking?
We assume that the fermion flavor $N$ is larger than the critical
value $N_c$, so chiral symmetry is not broken. Even though the
fermions are massless, we will argue that fermion confinement can be
induced by an alternative mechanism.

In any realistic many-particle systems, the fermion damping effect
is always important. Technically, the fermion damping can be
represented by including an imaginary self-energy $i\Gamma$ to the
fermion propagator \cite{Altland}. In practice, the fermion damping
rate $\Gamma$ determines most of the low-energy transport properties
of a many-particle system \cite{Altland}. To understand the meaning
of fermion damping, it is convenient to work in the Matsubara
formalism and write the fermion propagator as
\begin{eqnarray}\label{eq:propagator_Gamma}
S(i\omega_n,\mathbf{p}) = \frac{1}{i\omega_n\gamma_0 -
\mathbf{\gamma}\cdot \mathbf{p}},
\end{eqnarray}
where the frequency is $i\omega_n = i(2n+1)\pi/\beta$ with $\beta =
1/T$. In the presence of finite fermion damping, the fermion
frequency should be replaced by \cite{Altland}
\begin{equation}
i\omega_{n} \rightarrow i\omega_{n} + i\Gamma
\mathrm{sgn}(\omega_{n}).
\end{equation}
Making analytical continuation, $i\omega_{n} \rightarrow \omega +
i\delta$, we have the retarded fermion propagator
\begin{eqnarray}\label{eq:propagator_Gamma}
S_{\mathrm{ret}}(\omega,\mathbf{p}) = \frac{1}{(\omega +
i\Gamma)\gamma_{0} - \mathbf{\gamma}\cdot \mathbf{p}}.
\end{eqnarray}
After Fourier transformation, it becomes
\begin{equation}
S(t,\mathbf{r}) = e^{i\omega t - \Gamma t}e^{i\mathbf{p}\cdot
\mathbf{r}}
\end{equation}
in the real space-time. At a beginning time, $t = 0$, the fermion
stays in a plane wave state characterized by quantum numbers
$\omega$ and $\mathbf{p}$. Due to various interactions, this
specific fermion state damps as time $t$ increases, with $\Gamma$
measuring how rapidly the fermion state damps.

The fermion damping may be divided into two types. First of all, the
interaction with gauge field can give rise to fermion damping.
However, this kind of damping is caused by inelastic scattering and
thus is required by Pauli exclusion principle to vanish at zero
energy. In fact, we have carefully calculated the fermion damping
rate due to gauge field and found that it vanishes as $\propto
\omega^{1/2}$ at low energy \cite{WangLiu}. Besides gauge field, the
fermions can also be scattered by various static disorders
(impurities, defects, \emph{etc}.), which are usually unavoidable in
realistic many-particle systems. The scattering by static disorder
is elastic in the sense that no energy is transferred and thus can
lead to finite damping rate even at zero energy \cite{Altland}.
Comparing with the damping rate $\Gamma$ coming from elastic
scattering, the damping rate $\propto \omega^{1/2}$ from inelastic
scattering can be reasonably neglected. In this paper, we will use a
constant $\Gamma$ to describe the fermion damping effect and then
calculate the classical potential $V(\mathbf{r})$.

The fermion damping rate modifies the propagator to
\begin{eqnarray}\label{eq:propagator_full}
S(i\omega_{n},\mathbf{p}) = \frac{1}{(i\omega_{n} +
i\Gamma\mathrm{sgn}(\omega_{n}))\gamma_{0} - \mathbf{\gamma}\cdot
\mathbf{p}}
\end{eqnarray}
with energy $p_0$ being replaced by discrete frequency
$i\omega_{n}$. At finite temperature $T$, the polarization tensor is
\begin{eqnarray}
\Pi_{\mu \nu}(q_0,\mathbf{q}) &=& -\frac{\alpha}{\beta}
\sum_{\omega_n}\int\frac{d^2 \mathbf{k}}{(2\pi)^2}
\frac{\mathrm{Tr}[\gamma_\mu
k\!\!\!/\gamma_\nu(k\!\!\!/+q\!\!\!/)]}{k^2(k+q)^2} \nonumber \\
&=& \Pi_A(q_0,\mathbf{q})A_{\mu \nu}+\Pi_B(q_0,\mathbf{q})B_{\mu
\nu},
\end{eqnarray}
where
\begin{eqnarray}
A_{\mu\nu} &=& (\delta_{\mu 0} - \frac{q_\mu
q_0}{q^2})\frac{q^2}{\mathbf{q}^2}(\delta_{\nu0} - \frac{q_\nu
q_0}{q^2}), \\
B_{\mu\nu} &=& \delta_{\mu i}(\delta_{ij} - \frac{q_i
q_j}{\mathbf{q}^2})\delta_{j \nu},
\end{eqnarray}
which satisfy the relationship $A_{\mu\nu} + B_{\mu\nu} =
\delta_{\mu\nu} - \frac{q_\mu q_\nu}{q^2}$. Using these definitions,
we have
\begin{eqnarray}
\Pi_A &=& \frac{q^2}{\mathbf{q}^2}\Pi_{00}, \\
\Pi_B &=& \Pi_{ii} - \frac{q_0^2}{\mathbf{q}^2}\Pi_{00}.
\end{eqnarray}
Now the gauge field propagator can be recast in the form
\begin{eqnarray}
D_{\mu\nu}(q_0,\mathbf{q}) &=& \frac{A_{\mu \nu}}{q^2 + \Pi_A} +
\frac{B_{\mu\nu}}{q^2 +\Pi_B}.
\end{eqnarray}
It is hard to obtain the full analytical expression for
$\Pi_{\mu\nu}(q_0,\mathbf{q})$ when damping rate $\Gamma$ is finite.
However, from Eq.(\ref{potential}), we know that we only need to
know the expression of $\Pi_{\mu\nu}(0,\mathbf{q})$. When $q_0 = 0$,
it is possible to evaluate the polarization functions following the
techniques outlined in previous papers \cite{Dorey, Liu09}. After
tedious but straightforward calculation, we finally have
\begin{widetext}

\begin{eqnarray}
\Pi_{00}(\mathbf{q}) &=& \frac{2\alpha}{\pi^2}\int^1_0 dx \frac{y^2
- x(1-x)\mathbf{q}^2}{y}\mathrm{Im}\left[\psi\left(\frac{1}{2} +
\frac{\Gamma}{2\pi T}+i\frac{y}{2\pi T}\right)\right]^\Lambda_
{\sqrt{x(1-x)\mathbf{q}^2}} \nonumber \\
&& -\frac{2\alpha}{\pi^2}\int^1_0 dx
\int^\Lambda_{\sqrt{x(1-x)\mathbf{q}^2}}dy
\mathrm{Im}\left[\psi\left(\frac{1}{2}+\frac{\Gamma}{2\pi T} +
i\frac{y}{2\pi T}\right)\right], \nonumber \\
\Pi_{ij}(\mathbf{q}) &=& \frac{2\alpha}{\pi^2}\int^1_0 dx
\sqrt{x(1-x)\mathbf{q}^2}\left(\delta_{ij} - \frac{q_i
q_j}{\mathbf{q}^2}\right)\mathrm{Im}\left[\psi\left(\frac{1}{2} +
\frac{\Gamma}{2\pi T} + i\frac{\sqrt{x(1-x)\mathbf{q}^2}}{2\pi
T}\right)\right].
\end{eqnarray}
\end{widetext}
In the $\Gamma = 0$ limit, the integration over parameter $y$ is
convergent and the polarization function takes the well-known
expression $\Pi(q) \propto \alpha/q$. When $\Gamma \neq 0$, the
above integration is divergent in the region $y \rightarrow
+\infty$, so an ultraviolet cutoff $\Lambda$ has to be introduced.
The $\psi$ function is very complicated and in general can not be
analytically integrated. However, taking the zero temperature limit,
we can invoke the approximation
\begin{eqnarray}
\lim_{T \rightarrow 0} \mathrm{Im}\left[\psi\left(\frac{1}{2} +
\frac{\Gamma}{2\pi T} + i\frac{y}{2\pi T}\right)\right] =
\arctan\frac{y}{\Gamma}.
\end{eqnarray}
Then the integration over parameter $y$ can be carried out. Taking
advantage of the simplification $q_0 = 0$, we obtain the temporal
component of polarization function
\begin{eqnarray}
\Pi_A(\mathbf{q}) &=& \frac{2\alpha}{\pi^2}\int^1_0
dx\sqrt{x(1-x)\mathbf{q}^2}
\arctan\frac{\sqrt{x(1-x)\mathbf{q}^2}}{\Gamma} \nonumber \\
&& + \frac{2\alpha}{\pi^2} \Gamma\ln\frac{\Lambda}{\Gamma} -
\frac{\alpha}{\pi^2}\int^1_0 dx\Gamma
\ln\left[1+\frac{x(1-x)\mathbf{q}^2}{\Gamma^2}\right], \nonumber \\
\end{eqnarray}
and the spatial component of polarization function
\begin{eqnarray} \label{pi_B}
\Pi_B(\mathbf{q}) &=& \frac{2\alpha }{\pi^2}\int^1_0 dx
\sqrt{x(1-x)\mathbf{q}^2}
\arctan\frac{\sqrt{x(1-x)\mathbf{q}^2}}{\Gamma}. \nonumber \\
\end{eqnarray}

When $\mathbf{q} \rightarrow 0$, the temporal polarization function
is
\begin{eqnarray}
\Pi_A(0) = \frac{2\alpha}{\pi^2} \Gamma\ln\frac{\Lambda}{\Gamma}.
\end{eqnarray}
This expression indicates that the temporal component of gauge
interaction is statically screened by developing an effective mass.
It is necessary to emphasize that the effective mass $\propto
\Gamma\ln\frac{\Lambda}{\Gamma}$ is precisely the density of states
of massless fermions at the Fermi energy produced by disorder
scattering \cite{Gorkov, Durst}. This reminds us of the static
screening of Coulomb interaction, which corresponds to the temporal
component of electromagnetic field, in non-relativistic metals,
where the Debye screening factor is also proportional to the
electron density of states. When momentum $\mathbf{q}$ is much
smaller than $\Gamma$, the spatial component of polarization
function has the expression
\begin{eqnarray}
\Pi_B(\mathbf{q}) = \frac{\alpha}{3\pi^2 \Gamma}\mathbf{q}^2.
\end{eqnarray}
Unlike the temporal component, the spatial component of gauge
interaction remains long-ranged since the gauge invariance requires
that no mass of gauge boson can be generated unless the local gauge
invariance is spontaneously broken (Higgs mechanism). As a
consequence, the massive temporal component of gauge field can be
simply neglected and it is only necessary to study the spatial
component.

It is important to notice that $\Pi_B(\mathbf{q})$ is proportional
to $\mathbf{q}^2$, just like the free Maxwell term. Therefore, the
classical potential between fermions becomes
\begin{eqnarray}
V(\mathbf{r}) \propto \frac{1}{1+\frac{\alpha}{3\pi^2 \Gamma}}\ln r
+ O(\frac{1}{r}).
\end{eqnarray}
It is now clear that a finite damping rate induces weak confinement
of massless fermions. If we alternatively take the $\Gamma = 0$
limit in Eq.(\ref{pi_B}), then the polarization function is
proportional to $|\mathbf{q}|$ and therefore does not lead to
fermion confinement, as already pointed out by Burden \emph{et}
\emph{al.} \cite{Burden} and Maris \cite{Maris}.

We now know that both fermion damping rate $\Gamma$ and fermion mass
$m$ can give rise to a logarithmic confining potential between
fermions. It seems that they have the same effects on the behavior
of confinement. However, we have to emphasize that these two
quantities are physically quite different. When the fermion acquires
a finite mass $m$, there appears an effective $\propto
m\bar{\psi}\psi$ term in the Lagrangian. This mass term breaks the
chiral symmetry. However, the fermion damping rate $\Gamma$ enters
the fermion propagator by replacing fermion energy $p_0$ by $p_0 +
i\Gamma$. This is equivalent to introducing a term $\propto
i\Gamma\bar{\psi}\gamma_0 \psi$ to the Lagrangian. Apparently, this
term only leads to fermion damping but does not break the chiral
symmetry.

In summary, in this paper we studied the effect of fermion damping
on the classical potential between fermions in QED$_3$. By
analytically calculating the temporal and spatial components of the
vacuum polarization function, we found that a finite fermion damping
rate leads to logarithmic confinement even when the fermions remain
massless and the chiral symmetry is not broken. The massless
fermions exhibit completely different low-energy behaviors in
confined and deconfined phases, so the results obtained in this
paper may help to gain more understanding on several realistic
condensed matter systems, including high temperature superconductor
and certain spin liquid.

G.Z.L. is grateful to the financial support from the Project
Sponsored by the Overseas Academic Training Funds (OATF), University
of Science and Technology of China (USTC).

\end{document}